\documentclass[11pt]{iopart}

\usepackage{amsbsy,amssymb,mathrsfs}

\newcommand{\pl}{\partial}
\newcommand{\Ga}{\Gamma}
\newcommand{\al}{\alpha}
\newcommand{\be}{\beta}

\newcommand{\de}{\delta}
\newcommand{\m}{\mu}
\newcommand{\n}{\nu}

\newcommand{\eps}{\epsilon}

\newcommand{\Si}{\Sigma}

\newcommand{\mc}{\mathcal}

\newcommand{\ce}{\mc{E}}
\newcommand{\cm}{\mc{M}}

\newcommand{\cl}{\mc{L}}

\newcommand{\ca}{\mc{A}}

\newcommand{\ch}{\mc{H}}
\newcommand{\cq}{\mc{Q}}
\newcommand{\cg}{\mc{G}}
\newcommand{\cf}{\mc{F}}
\newcommand{\en}{n}
\newcommand{\nn}{\nonumber}
\newcommand{\ex}{\mc{C}}

\newcommand{\sch}{{\mbox{s}}}
\newcommand{\veh}{{\mbox{v}}}
\newcommand{\teh}{{\mbox{\tiny{T}}}}

\newcommand{\gee}{T}
\newcommand{\dphi}{\Gamma}

\begin{document}

\title[1+1+2 gravitational perturbations on LRS class II space-times]{1+1+2 gravitational perturbations on LRS class II space-times: GEM scalar harmonic amplitudes}

\author{R. B. Burston}

\address{Max Planck Institute for Solar System Research,
37191 Katlenburg-Lindau, Germany}
\eads{\mailto{burston@mps.mpg.de}}

\begin{abstract}
This is the third in a series of papers which considers first-order gauge-invariant and covariant gravitational perturbations to {\it locally rotationally symmetric} (LRS) class II space-times. In this paper we complete our analysis of the first-order {\it gravito-electromagnetic} (GEM) system by showing how to derive three decoupled equations governing the GEM scalar fields. One of these is for the gravito-magnetic scalar, whereas another two arise from the 2-gradient of the gravito-electric scalar.
\end{abstract}

\maketitle

\section{Introduction}

Paper I in this series \cite{Burston2007GEMT} uses similar methods that we used to decouple electromagnetic (EM) perturbations \cite{Burston2007EMVH,Burston2007EMBP} to show how to decouple the {\it gravito-electromagnetic} (GEM) 2-tensor harmonic amplitudes. This was followed by Paper II which showed that further decoupling is achieved when the complex GEM 2-vector is combined with the shear 2-tensors describing the 2/3-sheets \cite{Burston2007GEMV}. Summarily, from Paper's I and II we have found eight specific combinations that decouple which are
\begin{eqnarray}
\fl \mbox{Decoupled polar perturbations:     }   \Bigl\{ (\ce_\teh + \bar\ch_\teh),(\ce_\teh- \bar\ch_\teh) ,\nn\\
\fl  \Bigl[\Bigl( \ce_\veh+\frac 32\,\ce\, r\,\zeta_\teh\Bigr)-\Bigl(\bar \ch_\veh+\frac 32\,\ce\, r\,\Si_\teh\Bigr)\Bigr], 
 \Bigl[\Bigl(\ce_\veh+\frac 32\,\ce\, r\,\zeta_\teh\Bigr)+\Bigl(\bar \ch_\veh+\frac 32\,\ce\, r\,\Si_\teh\Bigr)\Bigr]\Bigr\} \\
\fl \mbox{Decoupled axial perturbations:     }\{(\ch_\teh  + \bar\ce_\teh),(\ch_\teh- \bar\ce_\teh) ,\nn\\
 \fl \Bigl[\Bigl(\ch_\veh+\frac 32\,\ce\,r\,\bar \Si_\teh\Bigr) +\Bigl( \bar\ce_\veh -\frac 32\,\ce\,r\,\bar \zeta_\teh\Bigr)\Bigr] , \Bigl[\Bigl(\ch_\veh +\frac 32\,\ce\,r\,\bar \Si_\teh\Bigr)- \Bigl(\bar\ce_\veh -\frac 32\,\ce\,r\,\bar \zeta_\teh\Bigr)\Bigr]\} .
\end{eqnarray}
This paper concentrates on decoupling the quantities governing the 1+1+2 GEM scalars. In Section \ref{newgemsys}, we provide a brief review of the complex GEM system we expressed in Paper II and its dependent variables.  Following this in Section \eref{decoudsvc} we show how decouple a combination of the 2-gradient of the gravito-electric scalar with other 1+1+2 quantities and the gravito-magnetic scalar.

\section{Complex 1+1+2 GEM system}\label{newgemsys}

The 1+1+2 complex GEM system was first expressed in Paper I and subsequently new dependent variables were chosen in Paper II and the new GEM system becomes
\begin{eqnarray}
\fl \Bigl(\cl_u-\frac 52\,\Si+\frac 53\,\theta\Bigr) \Xi_\m +\rmi\,{\epsilon_\m}^\al \Big(\cl_\en +\frac 52\,\phi\Bigr) \Xi_\al +\rmi\,  {\eps_\m}^\al(\de^2-K-3\,\ce)  \Psi_\al  \nn\\
-3\,\ce\,\Bigl[\Bigl(\Si-\frac 23\,\theta\Bigr)\dphi_\m+\rmi\,\phi\,{\eps_\m}^\al \dphi_\al \Bigr]  = \gee_\m \label{eqdxi},\\
\fl (\cl_\en+2\,\phi)\Psi_{\bar\m} -\rmi\,\frac 32\,\Si\,{\epsilon_\m}^\al \, \Psi_\al -\frac 12\,\Xi_\m+ (\de^2+K+3\,\ce) \dphi_\m = \tilde \cg_\m,\label{lenpsi}\\
\fl\Bigl(\cl_u-2\Si+\frac 43\theta\Bigr) \Psi_\m -\rmi {\eps_\m}^\al\Bigl(\ca-\frac 12\phi\Bigr)\Psi_\al +\rmi\frac 12{\eps_\m}^\al \Xi_\al  + \rmi(\de^2+K+3\ce) {\eps_\m}^\al \dphi_\al = \tilde\cf_\m, \label{lupsi}\\
\fl \Bigl(\cl_u +\frac32\,\Si+\theta\Bigr)\dphi_{\bar\m}
-\rmi\,{\epsilon_\m}^\al (\cl_\en+2\,\ca +\frac 12\,\phi) \dphi_\al+ \rmi\,\frac 12\,{\epsilon_\m}^\al \,\Psi_\al=\de^\al\cf_{\m\al}. \label{divphi}
\end{eqnarray}
where the new variables were defined
\begin{eqnarray}
\Xi_\m : = (\de^2+K)\ex_\m -3\,\ce\,\Bigl[ \phi\, \de^\al \zeta_{\m\al} +\Bigl(\Si-\frac 23\,\theta\Bigr) \,\de^\al \Si_{\m\al} +\de ^\al \Pi_{\m\al} \Bigr], \label{deffocu}\\
\Psi_\m:=(\de^2+K) \Phi_\m +\rmi\, 3\,\ce\, {\eps_\m}^\al \de^\be \Lambda_{\al\be},\\
\Ga_\m :=\de^\al \Phi_{\m\al}.
\end{eqnarray}

\section{Decoupling $\Xi_\m$ and its harmonic amplitudes}\label{decoudsvc}

In order to decouple $\Xi_\m$, we construct higher-order derivatives by first taking the Lie derivative of \eref{eqdxi} with respect to $u^\m$ and after much manipulation, we find
\begin{eqnarray}
\Bigl[\Bigl(\cl_u-4\,\Si+\frac {11}3\,\theta\Bigr)\cl_u-(\cl_\en +\ca+5\,\phi) \cl_\en - V \Bigr]\, \Xi_{\bar\m}= \cm_\m  \label{rwedA},
\end{eqnarray}
where the potential and energy-momentum source are
\begin{eqnarray}
V:= \de^2+19\, K+12\,\ce,\\
\fl \cm_\m := \Bigl(\cl_u -\frac 32\,\Si +2\,\theta\Bigr) \gee_\m -\rmi \, {\eps_\m}^\al \Bigl(\cl_\en +\ca+\frac 52\, \phi\Bigr) \gee_\al +3\,\ce \,\Bigl(\Si-\frac 23\,\theta\Bigr) \de^\al \cf_{\m\al}  \nn\\
 +\rmi \,3\,\ce\, \phi\,{\eps_\m}^\al \de^\be \cf_{\al\be}+(\de^2-K-3\,\ca) (\de^2+K)( \cg_\m-\rmi\,{\eps_\m}^\al \cf_\al).
\end{eqnarray}
This clearly demonstrates the decoupling of $\Xi_\m$ from the remaining first-order quantities. Furthermore, the differential operator in \eref{rwedA} is real and thus it is possible to consider the real and imaginary parts separately.

\subsection{Harmonic expansion}

The dependent variable $\Xi_\m$ and the energy-momentum source $\cm_\m$ are expanded into 2-vector harmonics according to
\begin{eqnarray}
\Xi_\m = \Xi_\veh\, Q_\m + \bar \Xi_\veh \,\bar Q_\m\qquad \mbox{and}\qquad \cm_\m = \cm_\veh\, Q_\m + \bar \cm_\veh \,\bar Q_\m.
\end{eqnarray}
Therefore, \eref{rwedA} naturally decouples into
\begin{eqnarray}
 \Bigl[\Bigl(\cl_u -5\,\Si+\frac{13}{3}\,\theta\Bigr) \cl_u -(\cl_\en +6\,\phi+\ca) \cl_\en - \check V \Bigr] \Xi_\veh = \cm_\veh ,\label{rw1}\\
 \Bigl[\Bigl(\cl_u -5\,\Si+\frac{13}{3}\,\theta\Bigr) \cl_u -(\cl_\en +6\,\phi+\ca) \cl_\en - \check V \Bigr] \bar \Xi_\veh =\bar  \cm_\veh .\label{rw2}
\end{eqnarray}
and the new potential has been defined
\begin{eqnarray}
\check V := -\frac{k^2}{r^2}  +30\,K+21\,\ce.
\end{eqnarray}
Now since the differential operators in \eref{rw1}-\eref{rw2} are purely real, the real and imaginary parts may be considered separately. It is of interest to see how the harmonic amplitudes of $\Xi_\m$ are related back to the harmonic amplitudes of the GEM scalars and other 1+1+2 quantities. By using the definition \eref{deffocu} it can be shown that
\begin{eqnarray}
\Re[\Xi_V] =r\,p\,\Bigl\{ X_V -\frac 32\,\ce\,r\,\Bigl[\phi\, \zeta_\teh +\Bigl(\Si-\frac 23\,\theta\Bigr)\,\Si_\teh + \Pi_\teh\Bigr]\Bigr\},\\
\Re[\bar \Xi_V] =r\,p\, \Bigl\{ \bar X_V +\frac 32\,\ce\,r\,\Bigl[\phi\,\bar  \zeta_\teh +\Bigl(\Si-\frac 23\,\theta\Bigr)\,\bar \Si_\teh +\bar  \Pi_\teh\Bigr]\Bigr\},\\
\Im[\Xi_\veh] =p\, \ch_\sch,\\
 \Im[ \bar \Xi_\veh] =0,
\end{eqnarray}
where
\begin{eqnarray}
p := \frac 1r\,\Bigl(2\, K-\frac{k^2}{r^2}\Bigr).
\end{eqnarray}
Therefore, there are only 3 equations here. One of them governs the gravito-magnetic scalar $\ch_\sch$ whereas the remaining two govern the 2-gradient of the gravito-electric scalar combined with the 2-divergence of the 2-tensors for the shear of the 2/3-sheets and the anisotropic stress 2-tensor.

The factors $p$ and $r$ arise from a harmonic expansion of the 2-Laplacian in terms of harmonics and they can be differentiated and factorized if desired. For example, the decouple equation governing $\ch_\sch$ becomes
\begin{eqnarray}
p\,\Bigl[ \Bigl(\cl_u -2\,\Si+\frac 73 \,\theta\Bigr) \cl_u -(\cl_\en +\ca+3\,\phi)\cl_\en   - V_\ch \Bigr] \ch_\sch =  \Im[\cm_V]
\end{eqnarray}
where
\begin{eqnarray}
V_\ch := -\frac {k^2}{r^2} +\frac 32\, \phi^2-6\,\ce -\frac 32\,\Bigl(\Si-\frac 23\,\theta\Bigr)\Bigl(\Si+\frac 13\,\theta\Bigr).
\end{eqnarray}

Thus the three quantities which each decouple are
\begin{eqnarray}
\fl \mbox{Decoupled polar perturbations:     }   \Bigl\{ X_V -\frac 32\,\ce\,r\,\Bigl[\phi\, \zeta_\teh +\Bigl(\Si-\frac 23\,\theta\Bigr)\,\Si_\teh + \Pi_\teh\Bigr]\Bigr\} ,\\
\fl \mbox{Decoupled axial perturbations:     }\Bigl\{\bar X_V +\frac 32\,\ce\,r\,\Bigl[\phi\,\bar  \zeta_\teh +\Bigl(\Si-\frac 23\,\theta\Bigr)\,\bar \Si_\teh +\bar  \Pi_\teh\Bigr],\, \ch_\sch\Bigr\} .
\end{eqnarray}

\subsection{Reduction to the covariant Schwarzschild space-times}

We now consider the covariant Schwarzschild space-time such that these results may be related to previous work.  The only non-vanishing LRS class II scalars for this case are $(\ca,\phi,\ce)$ for which  we now have
\begin{eqnarray}
\Re[\Xi_\m] : = (\de^2+K)X_\m -3\,\ce\, \phi\, \de^\al \zeta_{\m\al} ,\label{redxim}\\
\Im[\Xi_\m] := (\de^2+K)\de_\m \ch_\sch.
\end{eqnarray}
Clarkson and Barret  \cite{Clarkson2003} showed how to derive a Regge-Wheeler (RW) \cite{Regge1957} 2-tensor $\mc{W}_{\m\n}$ in this case and one can show that this tensor is related to \eref{redxim} according to
\begin{eqnarray}
-\frac{6\,\ce}{r^2} \de^\al \mc{W}_{\m\al}= \Re[\Xi_\m] .
\end{eqnarray}
The imaginary part is related to work from the Newman-Penrose formalism in \cite{Price1972}. The decoupled equation we have derived for $\ch_\sch$ reduces to
\begin{eqnarray}
p\,\Bigl[ \cl_u \cl_u -(\cl_\en +\ca+3\,\phi)\cl_\en   +\frac {\ell(\ell+1)}{r^2} -\frac 32\, \phi^2+6\,\ce \Bigr] \ch_\sch =  0.
\end{eqnarray}
A scaling is introduced to facilitate comparison with the work of Price \cite{Price1972}, $\ch_\sch^\prime := r^3\, \ch_\sch$,
\begin{eqnarray}
\frac p{r^3}\,\left(\cl_u \cl_u  -(\cl_\en +\ca) \cl_\en +\frac{\ell(\ell+1)}{r^2} +3\,\ce\right)\ch_\sch^\prime=0
\end{eqnarray}
 We now expand everything in terms of coordinates, transform to the ``tortoise" coordinate, $r\star$, and omit the common factor to reveal,
\begin{eqnarray}
\left( \pl_t^2 - \pl_{r\star}^2 + V_{RW}  \right) \ch_\sch^\prime= 0,
\end{eqnarray}
where the standard RW potential is
\begin{eqnarray}
V_{RW} := \left(1-\frac{2\, M}r\right)\left( \frac{\ell(\ell+1)}{r^2}-\frac{6\, M}{r^3}\right),
\end{eqnarray}
and $M$ is the mass. This corresponds to Price's result that $ r^3\,\Im [\Psi_0] $ is a RW quantity \cite{Price1972} and $\Psi_0$ is a Newman-Penrose scalar.

\section{Summary}

We have provided a comprehensive analysis of the first-order 1+1+2 complex GEM system spanning three papers. This paper delivered the final three decoupled quantities and for completeness, we write the total of 11 decoupled quantities that we have found,

\begin{eqnarray}
\fl \mbox{Decoupled polar perturbations:     }   \Bigl\{  \nn\\
  (\ce_\teh + \bar\ch_\teh),\nn\\
 (\ce_\teh- \bar\ch_\teh) ,\nn\\
  \Bigl[\Bigl( \ce_\veh+\frac 32\,\ce\, r\,\zeta_\teh\Bigr)-\Bigl(\bar \ch_\veh+\frac 32\,\ce\, r\,\Si_\teh\Bigr)\Bigr],  \nn\\
  \Bigl[\Bigl(\ce_\veh+\frac 32\,\ce\, r\,\zeta_\teh\Bigr)+\Bigl(\bar \ch_\veh+\frac 32\,\ce\, r\,\Si_\teh\Bigr)\Bigr] , \nn\\
  X_V -\frac 32\,\ce\,r\,\Bigl[\phi\, \zeta_\teh +\Bigl(\Si-\frac 23\,\theta\Bigr)\,\Si_\teh + \Pi_\teh\Bigr]\Bigr\} \\
\fl \mbox{Decoupled axial perturbations:     }\Bigl\{\nn\\
 (\ch_\teh  + \bar\ce_\teh), \nn\\
 (\ch_\teh- \bar\ce_\teh) ,\nn\\
  \Bigl[\Bigl(\ch_\veh+\frac 32\,\ce\,r\,\bar \Si_\teh\Bigr) +\Bigl( \bar\ce_\veh -\frac 32\,\ce\,r\,\bar \zeta_\teh\Bigr)\Bigr] , \nn\\
   \Bigl[\Bigl(\ch_\veh +\frac 32\,\ce\,r\,\bar \Si_\teh\Bigr)- \Bigl(\bar\ce_\veh -\frac 32\,\ce\,r\,\bar \zeta_\teh\Bigr)\Bigr] ,\nn\\
  \bar X_V +\frac 32\,\ce\,r\,\Bigl[\phi\,\bar  \zeta_\teh +\Bigl(\Si-\frac 23\,\theta\Bigr)\,\bar \Si_\teh +\bar  \Pi_\teh\Bigr], \nn\\
  \ch_\sch\Bigr\} .
\end{eqnarray}


\section*{References}

\begin{thebibliography}{25}


\bibitem{Burston2007GEMT} Burston R B 2007 {\it 1+1+2 gravitational perturbations on LRS class II space-times: GEM vector harmonic amplitudes} (gr-qc)


\bibitem{Burston2007EMVH} Burston R B 2007 {\it 1+1+2 electromagnetic perturbations on LRS class II space-times: Decoupling vector and scalar harmonic amplitudes} submitted to \CQG

\bibitem{Burston2007EMBP} Burston R B and Lun A W C 2007 {\it 1+1+2 electromagnetic perturbations on LRS space-times: Regge-Wheeler and Bardeen-Press equations} submitted to \CQG

\bibitem{Burston2007GEMV} Burston R B 2007 {\it 1+1+2 gravitational perturbations on LRS class II space-times: Decoupling GEM tensor harmonic amplitudes} submitted to \CQG

\bibitem{Clarkson2003} Clarkson C and Barrett R 2003 \CQG {\bf 20} 3855-84

\bibitem{Regge1957} Regge T and Wheeler J 1957 \PR {\bf 108} 1063

\bibitem{Price1972} Price R H (1972) {\it Phys. Rev. D} {\bf 5} 2439-54


\end{thebibliography}
\end{document}